\documentclass[12pt]{iopart}

\usepackage{iopams}
\usepackage{graphicx}
\newcommand \eq [1]{Eq.~(\ref{#1})}

\begin{document}

\title[]{Asymptotic solution for first and second order integro-differential equations}

\author{Mauro Bologna}

\address{Instituto de Alta Investigaci\'{o}n, Universidad de
  Tarapac\'{a}-Casilla 7-D Arica, Chile}
\ead{mauroh69@libero.it}
\begin{abstract}
This paper addresses the problem of finding an asymptotic solution
for first and second order integro-differential equations
containing an arbitrary kernel, by evaluating the corresponding
inverse Laplace and Fourier transforms. The aim of the paper is to
go beyond the tauberian theorem in the case of
integral-differential equations which are widely used by the
scientific community. The results are applied to the convolute
form of the Lindblad equation setting generic conditions on the
kernel in such a way as to generate a positive definite density
matrix, and show that the structure of the eigenvalues of the
correspondent liouvillian operator plays a crucial role in
determining the positivity of the density matrix.

\end{abstract}

\submitto{\JPA}
\maketitle

\section{Introduction}\label{secintro}
The Laplace and Fourier transforms are powerful tools widely used
in scientific fields such as mathematics, physics, biology and
chemistry. These transforms are often applied to linear partial
differential equations and integro-differential equations to
eliminate time and space dependence. The analytical solutions thus
obtained need to be inverted to the time and space domain
(see~\cite{bell} for a monograph on the Laplace transform).
Santos~\cite{santos} found a procedure for an analytical inversion
of the Laplace transform reducing the inversion formula to an
integration on the interval $[0,\infty)$. While literature on the
numerical inversion of the Laplace transform is rich
(see~\cite{dav} for a review) the analytical inversion still rests
mostly on the tauberian theorem. This paper focuses on first and
second order integro-differential equations containing an
arbitrary kernel. It is virtually impossible to list a complete
bibliography on the topic and for this reason the reader is
referred to the following exemplary
papers~\cite{fox,paolo,grig,lee,lee2}. In these works the authors
discuss the first order integro-differential equation of the form

\begin{equation}\label{lee_grig1}
 \frac{ d}{dt}F(t)-zF(t)=\int_{0}^{t}K(t-t')F(t')dt'
\end{equation}
or, after applying the Laplace transform, the equivalent equation

\begin{equation}\label{lee_grig2}
 \hat{F}(u)=\frac{F(0)}{u-z-\hat{K}(u)}
\end{equation}
where, by definition~$\hat{F}(u)=\int_{0}^{\infty}\exp[-u
t]F(t)dt$. The above equation, \eq{lee_grig2}, represents a
typical form that leads the process for finding a solution of the
original problem stated by \eq{lee_grig1}. Once the transformed
function is obtained the inversion process is often a difficult
task. In this paper we shall consider the inversion of a Laplace
transform of the form

\begin{equation}\label{l1}
\hat{F}(u)= \frac{1}{P_n(u)+\varepsilon \hat{K}(u)}
\end{equation}
and analogously, of a Fourier transform of the form

\begin{equation}\label{f1}
\hat{F}(\omega)= \frac{1}{P_n(\omega)+\varepsilon \hat{K}(\omega)}
\end{equation}
where~$P_n(u)$ or~$P_n(\omega)$ is a~$n$-degree polynomial,
$\varepsilon$ is a parameter, and~$\hat{K}(u)$ or
$\hat{K}(\omega)$ is an arbitrary function. Without loss of
generality we shall consider the case of the Laplace transform.
With slight changes, the results may be applied to the Fourier
transform.

The main goal of this paper is to give a prescription to find an
asymptotic expression for the function~$F(t)$ in the
representation of the starting variable, typically the time domain
for the Laplace transform, or the space domain for the Fourier
transform. This problem is partially solved by the use of the
tauberian theorem but, as it is well known, the conditions for
correctly applying this theorem are quite strict. We shall focus
on the case of~$P_1(u)=u+a$ and~$P_2(u)=u^2\pm a^2$ giving
sufficient conditions on~$\hat{K}(u)$ that allow us to find an
approximate expression for either the inverse Fourier or the
Laplace transforms.

This work is organized as follows: In Sec.~\ref{sec2} we consider
a short review for the case when the function~$\hat{K}(u)$ is a
generic polynomial such that an analytical expression for the
asymptotic solution $F(t)$ is given when~$P_n(u)$ is both a first
and second degree polynomial. In Sec.~\ref{sec4} we adopt the
multi-scale approach to find an approximate solution for the case
when~$P_n(u)$ is a second degree polynomial and~$\hat{K}(u)$ is a
generic function. In Sec.~\ref{sec5} the problem is solved in a
less generic way but it is mathematically rigorous. An equation
for the asymptotic expression of~$F(t)$ generated by
\eq{lee_grig1} shall be found. Such an equation is independent of
the form of the kernel. This is why we can use the term
\textit{universality} to describe the asymptotic equation.
Finally, in Sec~\ref{sec6} the previous results are applied to the
case of the convoluted Lindblad equation~\cite{lin,gor} whereby we
discuss the sufficient conditions on the kernel to obtain a
positive definite quantum density matrix.

\section{Laplace transform containing polynomials}\label{sec2}
In this section we briefly examine the polynomial case, however
before exposing the main idea let us clarify a key point.
Intuitively, we could say that for~$\varepsilon\to 0$ the inverse
Laplace transform of Eq.~(\ref{l1}) can be evaluated at the poles
of the unperturbed polynomial. Moreover, we could try to apply the
tauberian theorem to "guess" the asymptotic solution. The
following example shows that, in general, the problem can be much
more complex. To illustrate the main idea let us consider the
following Laplace transform

\begin{equation}\label{polex1}
\hat{F}(u)= \frac{1}{u+1+\varepsilon \left(u^3+u^2\right)}.
\end{equation}
Naively we could say that for~$\varepsilon\to 0$,
then~$F(t)\approx \exp[-t]$. First, let us find the solution
neglecting~$u^3$ employing the tauberian theorem idea. The poles
can be evaluated analytically and its expressions are~$$ u_1=
\frac{-1-\sqrt{1-4 \varepsilon }}{2 \varepsilon }
,\,\,\,\,u_2=\frac{-1+\sqrt{1-4 \varepsilon }}{2 \varepsilon }.~$$
Consequently the solution is

\begin{equation}\label{app1}
 F(t)=\frac{2 e^{-\frac{t}{2 \varepsilon }} \sinh
 \left[\frac{t \sqrt{1-4 \varepsilon }}{2 \varepsilon
}\right]}{\sqrt{1-4 \varepsilon }}\approx
-\exp\left[\frac{-t}{\varepsilon}+t\right]+\exp[-(1+\varepsilon)t]\approx
\exp[- t]
\end{equation}
which seems to support the idea that for~$\varepsilon\to 0$, then
$F(t)\approx \exp[-t]$. If we evaluate the exact poles
$$ u_1= -1,\,\,\,\, u_2 -\frac{\imath}{\sqrt{\varepsilon }},\,\,\,\,
u_3=\frac{\imath}{\sqrt{\varepsilon }},$$ the exact solution is

\begin{equation}\label{app2}
F(t)=\frac{e^{-t}}{1+\varepsilon
}+\frac{-\cos\left[\frac{t}{\sqrt{\varepsilon
}}\right]+\sqrt{\varepsilon } \sin\left[\frac{t}{\sqrt{\varepsilon
}}\right]}{1+\varepsilon }.
\end{equation}
Note that the limit for~$\varepsilon\to 0$ of solution
(\ref{app2}) does not exist. The previous example clarifies an
important point. In general, for~$\varepsilon \to 0$, it is not
correct to invert the Laplace transform evaluating the approximate
poles of the unperturbed polynomial [see \eq{app1}]. The reason
why the expansion in power of the variable~$u$ fails in the
polynomial case is because by neglecting higher powers we are
neglecting poles that are divergent for~$\varepsilon \to 0$.

We now consider a second degree polynomial for~$P_n(u)$,
namely~$P_2(u)$, and at the end of this section we shall consider
a first degree polynomial,~$P_1(u)$. Without loss of generality,
we shall focus on the case~$P_2(u)=u^2\pm a^2$. We start by
considering~$\hat{K}(u)=G_n(u)$ where~$G_n(u)=\sum_{k=0}^{n}a_k
u^k$ is a polynomial of~$n$ degree with~$n> 2$. If we want an
approximate solution of~$\varepsilon$ order we must evaluate all
the poles in addition to the two given by
$$\bar{u}_{1,2}=\pm \sqrt{- a^2-\varepsilon G_n(\sqrt{- a^2})}.$$
To fix the ideas, we select~$P_2(u)=u^2+ a^2$. Looking for a
scaling such that the term~$u^n$ is of the same order of~$u^2$, we
perform the transformation~$u\to \varepsilon^{\nu} U~$ so that we
have

\begin{equation}\label{f4_b}
\varepsilon^{2\nu}=\varepsilon^{n\nu+1}.
\end{equation}
Equating the exponents we find that~$\nu=-\frac{1}{n-2}$. Keeping
only the lowest order, we obtain the poles for the polynomial
equation

\begin{equation}\label{f4_c}
a_n U^n_0+ U^2_0=0.
\end{equation}
The solution can be easily found as

\begin{equation}\label{f4_d}
U_0(k) = \frac{1}{\mid a_n
\mid^{\frac{1}{n-2}}}\exp\left[\frac{\imath\phi}{n-2}+
\frac{2k\pi\imath}{n-2}\right],\,\,\,\,\,k=0,\cdots n-3
\end{equation}
with~$\phi=\arg(-a_n)$. \eq{f4_d} gives~$n-2$ solutions that,
combined with the two solutions given by~$$\bar{u}_{1,2}=\pm
\sqrt{- a^2-\varepsilon G_n(\sqrt{- a^2})},$$ complete the~$n$
solutions for the total polynomial. The next order for the~$n-2$
divergent solutions is given by:

\begin{equation}\label{f4_e}
U_1(k) = - \frac{a_{n-1}}{(n-2)a_{n}}.
\end{equation}
Using the residue theorem we find the inverse Laplace transform by
evaluating the integral

\begin{equation}\label{f4_ext}
F(t)= \frac{1}{2\pi\imath}\int\limits_{-\imath
\infty+\gamma}^{\imath \infty+\gamma}\frac{\exp[u
t]}{P_2(u)+\varepsilon G_n(u)} du
\end{equation}
at the poles given by

\begin{equation}\label{f4_f}
u_k =\frac{U_0(k)}{\varepsilon^{\frac{1}{n-2}}} -
\frac{a_{n-1}}{(n-2)a_{n}},\,\,k=0\cdots n-3,
\,\,\bar{u}_{1,2}=\pm \sqrt{- a^2-\varepsilon G_n(\sqrt{- a^2})}.
\end{equation}
The approximate expression for the function~$F(t)$ is

\begin{equation}\label{f5}
F(t)= \frac{\exp[\bar{u}_1 t] }{P_2'(\bar{u}_1)+\varepsilon
G_n'(\bar{u}_1)}+\frac{\exp[\bar{u}_2 t]
}{P_2'(\bar{u}_2)+\varepsilon
G_n'(\bar{u}_2)}+\sum\limits_{k=0}^{n-3}\frac{\exp[u_k t]
}{P_2'(u_k)+\varepsilon G_n'(u_k)}
\end{equation}
where~$P_2'(u)$ and~$ G_n'(u)$ are the derivatives of the
polynomials evaluated in~$u_k$ and~$\bar{u}_{1,2}$. Similarly, for
the poles of the first degree polynomial case, $P_1(u)=u+ a$, we
obtain the following expressions,
$$\bar{u}_{1}=-a-\varepsilon G_n(-a)$$
and~$$ u_k =\frac{1}{\mid a_n
\mid^{\frac{1}{n-1}}\varepsilon^{\frac{1}{n-1}}}\exp\left[\frac{\imath\phi}{n-1}+
\frac{2k\pi\imath}{n-1}\right]+\frac{a-\frac{a_{n-1}}{a_n}}{n-1}$$
for~$k=0,\cdots n-2$.

\section{Laplace transform containing a generic function }\label{sec4}

In this section we shall consider a Laplace transform containing a
generic function~$\hat{K}(u)$ that can be developed in the Taylor
series at the unperturbed poles given by the zeros of the
polynomial~$P_n(u)$. As shown in Sec.~\ref{sec2} we can not
neglect the higher terms of~$u$ powers. Nevertheless, the
asymptotic behavior of~$\hat{K}(u)$ gives us information about the
"effective" power of~$\hat{K}(u)$. We set the condition that
$\hat{K}(u)$ grows slower than~$P_n(u)$, more precisely

\begin{equation}\label{f2_b}
\lim_{u\to\infty} \frac{\hat{K}(u)}{P_n(u)}=0.
\end{equation}
It is important to stress that the limit is performed on the
positive real axis and not on the complex plane.

Next we choose the case of~$P_2(u)=u^2\pm a^2$ which results in a
typical expression, for example the calculus of Green's function
(see Eq.~(\ref{f1}), Ref.~\cite{fetter} for quantum applications).
Our goal is to evaluate the integrals

\begin{equation}\label{frint}
F(t)=\frac{1}{2\pi\imath}\int_{\Gamma}\frac{e^{u t}}{u^2\pm
a^2+\varepsilon  \hat{K}(u) }du,\,\,\,\,\,
F(x)=\int\limits_{-\infty}^{\infty}\frac{e^{\imath \omega x}}{
\omega^2\pm a^2+\varepsilon  \hat{K}(\omega) }\frac{d\omega}{2\pi}
\end{equation}
representing the inverse of the Laplace and Fourier transforms,
respectively. In general the straightforward expansion of the
integrand function in~$\varepsilon$ powers is not correct at
$t$-scale or~$x$-scale of the order of~$ 1/\varepsilon$. In other
words, the expansion in~$\varepsilon$ powers such as

\begin{equation}\label{f4}
 F(x)\approx\int\limits_{-\infty}^{\infty}\frac{e^{\imath
\omega x}}{ \omega^2\pm a^2}\frac{d\omega}{2\pi} -\varepsilon
\int\limits_{-\infty}^{\infty}\frac{e^{\imath \omega
x}\hat{K}(\omega) }{ (\omega^2\pm a^2)^2}\frac{d\omega}{2\pi}
\end{equation}
leads to an unsatisfactory approximation. It is also worthy to
stress that the second integral on the right side of \eq{f4} can
be as difficult to evaluate as the initial one, \eq{frint}.

In the time representation we write the equation for~$F(t)$ as

\begin{equation}\label{lapl2}
\left[\frac{d^2}{dt^2}+a^2+\varepsilon \hat{K}\left(
\frac{d}{dt}\right)\right]F(t)=0
\end{equation}
where the term~$\varepsilon \hat{K}\left(\frac{d}{dt}\right)$
descends directly per the hypothesis that~$\hat{K}(u)$ can be
developed at the unperturbed poles in the Taylor series. Since the
hypothesis states that~$\hat{K}(u)$ grows slower than $u^2$, we
deduce that when~$t\to 0$ then~$F(t)\approx t$. Using this
information we find that~$F(0)=0$ and~$F'(0)=1$. With regard to
\eq{f2_b} the "order" of the derivative given
by~$\hat{K}\left(\frac{d}{dt}\right)$ is smaller then the second
derivative so that we can consider the term~$\varepsilon
\hat{K}\left(\frac{d}{dt}\right)F(t)$ as a slow varying function
of~$t$. Applying the multiple scale technique~\cite{kev} we
write~$F(t)$ as

\begin{eqnarray}\label{f8}
F(t) &=&F(\tau_0,\tau_1,\tau_2,\cdots)=F_0(t)+\varepsilon
F_1(t)+\cdots,
\\
\label{f9} \frac{d}{dt}&=&\frac{\partial}{\partial
\tau_0}+\varepsilon\frac{\partial}{\partial \tau_1}+
\varepsilon^2\frac{\partial}{\partial \tau_2}+\cdots.
\end{eqnarray}
where, by definition~$\tau_0=t,\,\tau_1=\varepsilon
t,\,\tau_2=\varepsilon^2 t,\cdots$. The solution of \eq{lapl2} at
zero order in~$\varepsilon$ is

\begin{equation}\label{lapl4}
F_0(t)=A(\tau_1,\tau_2,\cdots)\exp[\imath a
\tau_0]+B(\tau_1,\tau_2,\cdots)\exp[-\imath a \tau_0].
\end{equation}
To determine the functions~$A(\tau_1,\tau_2,\cdots)$
and~$B(\tau_1,\tau_2,\cdots)$ we need the equation of the first
order in~$\varepsilon$,

\begin{equation}\label{lapl5}
\left[\frac{\partial^2}{\partial \tau_0^2}+a^2 \right]F_1(t)+
2\frac{\partial^2}{\partial \tau_0\partial
\tau_1}F_0(t)+\hat{K}\left( \frac{\partial}{\partial
\tau_0}\right)F_0(t)=0.
\end{equation}
The solution for~$A(\tau_1,\tau_2,\cdots)$
and~$B(\tau_1,\tau_2,\cdots)$ with the conditions~$F(0)=0$
and~$F'(0)=1$ is

\begin{equation}\label{lapl6}
F(t)\approx\frac{\exp\left[\imath a\left(1+\varepsilon
\frac{\hat{K}(\imath a)}{2 a ^2}\right)t\right]-\exp\left[-\imath
a\left(1+\varepsilon \frac{\hat{K}(-\imath a)}{2 a
^2}\right)t\right]}{2\imath a \left[1+\varepsilon
\left(\frac{\hat{K}(\imath a)}{4 a ^2} + \frac{\hat{K}(-\imath
a)}{4 a ^2}\right)\right]}.
\end{equation}
Now let us consider the following example
with~$\hat{K}(u)=u^\alpha$,

\begin{equation}\label{lapl7}
\hat{F}(u)=\frac{1}{u^2+1+\varepsilon u^\alpha}.
\end{equation}
According to \eq{lapl6}, the solution is approximated by

\begin{equation}\label{exl1}
F(t)\approx\frac{\exp\left[-\frac{\varepsilon}{2} t  \sin
\frac{\pi \alpha }{2} \right]\sin\left[t
\left(1+\frac{\varepsilon}{2}\cos \frac{\pi \alpha
}{2}\right)\right] }{1+\frac{\varepsilon}{2} \cos \frac{\pi \alpha
}{2} }.
\end{equation}
\begin{figure}[ht]
\begin{center}
\includegraphics[width=4.5cm, height=3.5cm,angle=0]{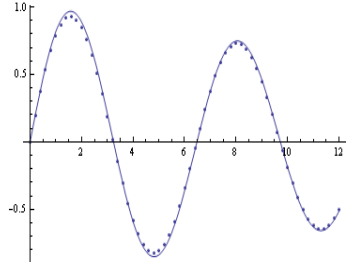}
\end{center}
\caption{The dots represent the numerical evaluation of the
inverse Laplace transform of \eq{lapl7}. The continuous line is
the approximate formula, \eq{exl1}. The values of the parameters
are~$\alpha=\sqrt{2}$ and $\varepsilon=0.1$.  \label{fig3}}
\end{figure}
The numerical check is shown in Fig.~\ref{fig3}.

\begin{figure}[ht]
\begin{center}
\begin{minipage}[t]{0.3\linewidth}
\centering
\includegraphics[width=3.5cm, height=3.5cm,angle=0]{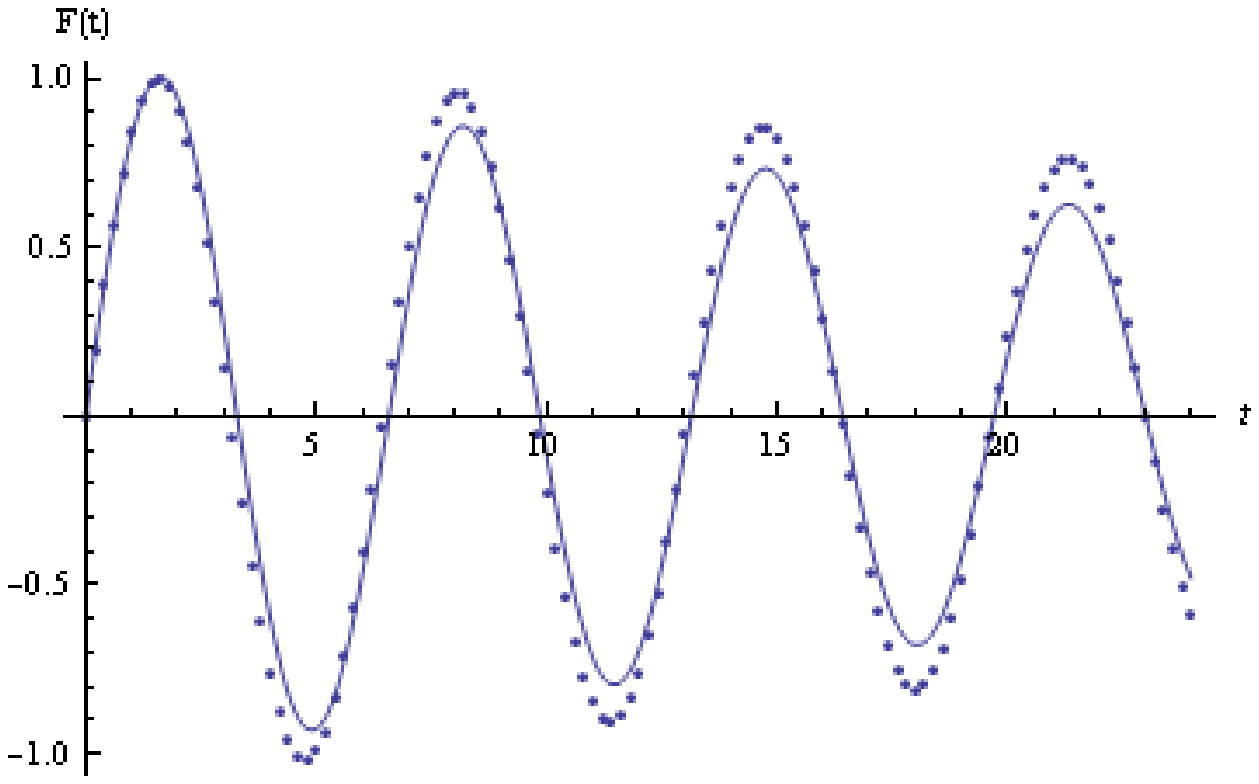}
\end{minipage}
\begin{minipage}[t]{.3\linewidth}
\centering
\includegraphics[width=3.5cm, height=3.5cm,angle=0]{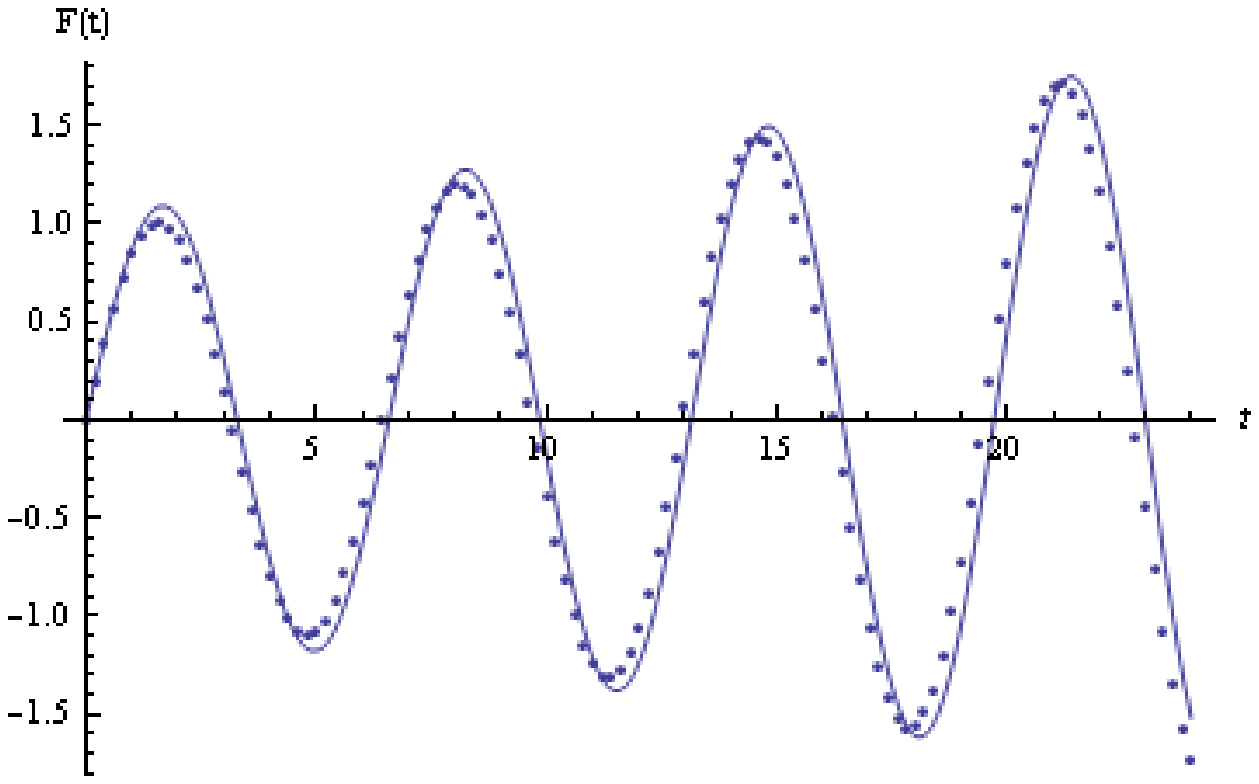}
\end{minipage}
\end{center}
\caption{\label{fig4} Left graphic: The dots represent the
numerical evaluation of the inverse Laplace transform of
\eq{ex12}. The continuous line is the approximate formula,
\eq{ex13}. The values of the parameters are~$b=\pi-0.5$
and~$\varepsilon=0.1$. Right graphic: The dots represent the
numerical evaluation of the inverse Laplace transform of
\eq{ex12}. The continuous line is the approximate formula,
\eq{ex13}. The values of the parameters are~$b=\pi+0.5$
and~$\varepsilon=0.1$. \label{fig5}}
\end{figure}

To emphasize the fact that the limit (\ref{f2_b}) is performed on
the positive real axis we consider~$\hat{K}(u)=\exp[-bu]$ as
kernel in the next example. In general the condition (\ref{f2_b})
is not satisfied for~$u\in C$, with~$C$ as the complex plane.
Writing~$\hat{F}(u)$ with the selected $\hat{K}(u)$ we have

\begin{equation}\label{ex12}
\hat{F}(u)=\frac{1}{u^2+1+\varepsilon  \exp[-bu]}.
\end{equation}
Applying \eq{lapl6}, we obtain

\begin{equation}\label{ex13}
F(t)= \frac{  \exp\left[\frac{1}{2} t \varepsilon \sin b
\right]\sin\left[t\left(1+\frac{1}{2} \varepsilon \cos
b\right)\right] }{1+\frac{\varepsilon }{2}\cos b}.
\end{equation}
Note that \eq{ex13} indicates that there is a critical
value,~$b=\pi$, for the parameter~$b$. This critical value
corresponds to an exponentially growing solution or to an
exponentially damped solution according to whether~$b<\pi$
or~$b>\pi$, respectively. The results are numerically checked in
Fig. \ref{fig4}.

\section{Universality of the asymptotic equation}\label{sec5}
In this section we will find an asymptotic expression for the
inverse Laplace transform of a function in the form (\ref{l1}). In
the previous section we focused our attention on the
case~$P_2(u)$, whereas here we will predominantly consider the
case~$P_1(u)$. This case has not yet been treated since the
multi-scale method essentially applies to second degree
polynomials. It shall be evident that, with slight changes, the
procedure can also be applied to the Fourier transform. We begin
with the following relatively simple case

\begin{equation}\label{fdp}
\hat{F}(u)= \frac{\hat{K}(u)}{u-z}
\end{equation}
where~$z$ is a complex parameter and~$\hat{K}(u)$ is the Laplace
transform of~$K(t)$. Making the hypothesis that the Laplace
transform is evaluated in~$z$, that is to say~$\hat{K}(z)$ exists,
we may write the solution of \eq{fdp} as

\begin{equation}\label{fdp_bis}
F(t)= \int_{0}^{t}K(t')\exp[z(t-t')]dt'=\exp[z t
]\int_{0}^{t}K(t')\exp[-zt']dt'
\end{equation}
and for~$t\to\infty$

\begin{equation}\label{fdp_ter}
F(t)= \exp[z t] \hat{K}(z).
\end{equation}
Following the same idea we now consider a more complex form
of~$\hat{F}(u)$. As said in Sec.~\ref{secintro} a Laplace
transform of the form

\begin{equation}\label{fdp1}
\hat{F}(u)= \frac{1}{u-z+\varepsilon \hat{K}(u)}
\end{equation}
has several physical applications. As we previously assumed the
function~$\hat{K}(u)$ has an inverse Laplace transform,~$K(t)$,
and~$\hat{K}(z)$ exists. Considering the time domain, \eq{fdp1} is
equivalent to the following equation
$$\frac{d}{dt} F(t)-z F(t)=-\varepsilon  \int_0^t
K(t-t')F(t')dt'.$$ Let us now examine \eq{fdp1} in more detail.
Developing its denominator in~$ \varepsilon$ power we have

\begin{equation}\label{fdp2}
\hat{F}(u)=
\frac{1}{u-z}\sum_{n=0}^{\infty}(-1)^n\frac{\varepsilon^{n}[
\hat{K}(u)]^{n}}{(u-z)^n}.
\end{equation}
Inverting the Laplace transform we obtain the following expression

\begin{eqnarray}\nonumber
F(t)&=&
\sum_{n=0}^{\infty}\frac{(-1)^n}{n!}\int_{0}^{t}\varepsilon^{n}
K_n(t')(t-t')^n\exp[z(t-t')]=\\
\label{fdp3} &=&
\sum_{n=0}^{\infty}\frac{(-1)^n}{n!}\varepsilon^{n}
\frac{\partial^n}{\partial z^n}\left[\exp[z t
]\int_{0}^{t}K_n(t')\exp[-z t' ]dt'\right]
\end{eqnarray}
where~$K_n(t)$ is the~$n$th convolution of~$K(t)$, and in the
recursive form, it is
$$K_n(t)=\int_{0}^{t}K(t-t')K_{n-1}(t')dt',\,\,\,\,\,\,K_0(t)=\delta(t).$$ Since we are
interested in the asymptotic behavior, we take the limit for
$t\to\infty$ resulting in
\begin{eqnarray}\nonumber
F(t)&\approx &\sum_{n=0}^{\infty}\frac{(-1)^n}{n!}\varepsilon^{n}
\frac{\partial^n}{\partial z}\left[\exp[z t
]\int_{0}^{\infty}K_n(t')\exp[-z t' ]dt'\right]
\\
\label{fdp4}
&=&\sum_{n=0}^{\infty}\frac{(-1)^n}{n!}\varepsilon^{n}
\frac{\partial^n}{\partial z^n}\left[\exp[z t] \hat{K}^n(z)
\right] \end{eqnarray} where, by definition, $ \hat{K}^n(z)\equiv
[ \hat{K}(z)]^n$. Applying simple algebra we obtain the following
asymptotic expression

\begin{equation}\label{fdp5}
F(t)=\exp[zt]\sum_{n=0}^{\infty}\frac{(-1)^n}{n!}\varepsilon^{n}\left[t+
\frac{\partial}{\partial z}\right]^n\left[\hat{K}^n(z)
\right]\equiv\exp[zt]\phi_1(\varepsilon,t,z)
\end{equation}
where the function~$\phi_1(\varepsilon,t,z)$ is defined as

\begin{equation}\label{vareps}\phi_1(\varepsilon,t,z)=
\sum_{n=0}^{\infty}\frac{(-1)^n}{n!} \varepsilon^{n}\left[t+
\frac{\partial }{\partial z}\right]^n\left[ \hat{K}^n(z)
\right].\end{equation} Note that the parameter~$\varepsilon$ is
not necessarily small. The only requirement is that \eq{vareps}
has to be a convergent series. The~$\phi_1(\varepsilon,t,z)$
function satisfies the following equation

\begin{equation}\label{fdp6}
\varepsilon\frac{\partial}{\partial\varepsilon
}\phi_1(\varepsilon,t,z)=\left[t\frac{\partial}{\partial t
}+\frac{\partial^2}{\partial z \partial
t}\right]\phi_1(\varepsilon,t,z).
\end{equation}
Changing the variables,~$\tau=\varepsilon t,\,\,v=\varepsilon$
and~$ w=z$, leads to a simplified version

\begin{equation}\label{fdp6_bis}
 \frac{\partial}{\partial v
}\phi_1(v,w,\tau)= \frac{\partial^2}{\partial w \partial
\tau}\phi_1(v,w,\tau).
\end{equation}
We can further transform \eq{fdp6_bis}.
Considering~$\phi_1(v,w,\tau)$ as the Laplace transform of a
function with respect to the variable $v$, and making the change
of variables~$w=x+y$ and~$\tau=x-y$, we may write \eq{fdp6_bis} as

\begin{equation}\label{fdp6_ter}
-\lambda \Phi_1(s,x,y)= \left[\frac{\partial^2}{\partial
x^2}-\frac{\partial^2}{\partial y^2}\right]\Phi_1(s,x,y)
\end{equation}
where, by definition

$$\Phi_1(s,x,y)\equiv
\frac{1}{2\pi\imath}\int\limits_{\gamma-\imath\infty}^{\gamma+\imath\infty}
\phi_1(v,x,y) \exp[sv]dv.$$ In \eq{fdp6_ter} we recognize the well
known Klein-Gordon equation. Using the complex change of
variables~$w=x+i y$ and~$\tau=x-iy$, we can rewrite \eq{fdp6_ter}
as

\begin{equation}\label{fdp6_iv}
-\lambda \Phi_1(s,x,y)= \left[\frac{\partial^2}{\partial
x^2}+\frac{\partial^2}{\partial y^2}\right]\Phi_1(s,x,y),
\end{equation}
that is the Helmholtz equation. The above equations;
Eqs.~(\ref{fdp6_bis}),~(\ref{fdp6_ter}) and~(\ref{fdp6_iv}), do
not contain the arbitrary function $\hat{K}(z)$. This fact implies
that all integro-differential equations generated by equations
like \eq{lee_grig1} are driven by the same asymptotic equation. In
this sense we can say that the asymptotic equation for the first
order integro-differential equations is universal.

To find a more manageable expression of $\phi_1(\varepsilon,t,z)$,
first we will rewrite it as

$$
\phi_1(\varepsilon,t,z)=\sum_{n=0}^{\infty}\frac{(-1)^n}{n!}\varepsilon^{n}t^n\left[1+
\frac{1}{t}\frac{\partial}{\partial z}\right]^n\left[ \hat{K}^n(z)
\right]=$$
$$=\sum_{n=0}^{\infty}\frac{(-1)^n}{n!}\varepsilon^{n}t^n\exp\left[n\log\left(1+
\frac{1}{t}\frac{\partial}{\partial z}\right)\right]\left[
\hat{K}(z) \right]^n\approx~$$
\begin{equation}\label{fdp7}
\approx\sum_{n=0}^{\infty}\frac{(-1)^n}{n!}\varepsilon^{n}t^n\exp\left[\frac{n}{t}
\frac{\partial}{\partial z}\right]\left[ \hat{K}(z)
\right]^n=\sum_{n=0}^{\infty}\frac{(-1)^n}{n!}\varepsilon^{n}t^n\left[
\hat{K}\left(\frac{n}{t}+z\right) \right]^n.
\end{equation}
Under the condition that~$\varepsilon$ is small enough to ensure a
fast convergence of the series, in such way that only the terms
$n\ll t$ contribute, we may write the following simplified
expression

\begin{equation}\label{fdp8}
\phi_1(\varepsilon,t,z)\approx\sum_{n=0}^{\infty}\frac{(-1)^n}{n!}
\varepsilon^{n}t^n\left[\hat{K}\left(z\right) \right]^n=\exp\left[
-\varepsilon \hat{K}\left(z\right)t \right].
\end{equation}
This implies

\begin{equation}\label{fdp9}
F(t)\approx\exp\left[\left(z-\varepsilon\hat{K}\left(z\right)\right)t
\right],
\end{equation}
and we can call this the \textit{naive solution} of the problem,
that is to say, the evaluation of the pole in \eq{fdp1} at the
unperturbed value $u=z$. We stress that in general the behavior
of~$\phi_1(\varepsilon,t,z)$ as function of time is more complex
than an exponential and the steps that lead from \eq{vareps} to
\eq{fdp8} confirm this statement. As an instructive example we may
evaluate \eq{vareps} in the case
where~$\hat{K}\left(z\right)=\exp[-\beta z]$. It is
straightforward to obtain for~$\phi_1(\varepsilon,t,z)$ the
expression

$$\phi_1(\varepsilon,t,z)=
\sum_{n=0}^{\infty}\frac{(-1)^n}{n!} \varepsilon^{n}\left[t-\beta
n\right]^n\exp\left[ -\beta n z\right].$$ The series is absolutely
convergent if the inequality~$\mid\!\!\varepsilon\beta \exp\left[
-\beta z +1\right]\!\!\mid<1$ is satisfied.

Following the lines previously expounded for the first degree
polynomial we shall find an expression for the case of a second
degree polynomial. Let us consider the following Laplace transform

\begin{equation}\label{sdp1}
\hat{F}(u)= \frac{1}{u^2\pm a^2+\varepsilon \hat{K}(u)}.
\end{equation}
For the sake of simplicity we assume that~$a$ is a real parameter
and we consider the plus sign in the denominator. We also need the
following result

\begin{eqnarray}\nonumber
\mathcal{L}^{-1}\left[\frac{1}{(u^2+a^2)^{n+1}}\right]&=&\frac{1}{a\left(2a\right)^{2n}n!}
\sum_{k=0}^{n}\frac{(n+k)!}{k!(n-k)!}(2at)^{n-k}\times\\\label{sdp2}
&\times &\sin\left[a t-\frac{\pi}{2}(n-k)\right].
\end{eqnarray}
Following the procedures set forth in this section we find the
exact expression for~$F(t)$

\begin{eqnarray}\nonumber
F(t)&=&\textrm{Im}\left[\sum_{n=0}^{\infty}\frac{(-\varepsilon)^{n}}{a\left(2a\right)^{2n}}
\frac{1}{n!}\sum_{k=0}^{n}\frac{(n+k)!}{k!(n-k)!}\times\right.
\\
\label{sdp3} &\times & \left. \int_{0}^{t}[-2\imath
a(t-t')]^{n-k}\exp[\imath a (t-t')]K_n(t')dt'\right]
\end{eqnarray}
and consequently the asymptotic expression

\[ F(t)\approx\textrm{Im} \left[\sum_{n=0}^{\infty}\frac{(2\imath a
\varepsilon)^{n}}{a\left(2a\right)^{2n}n!}
\sum_{k=0}^{n}\frac{(n+k)!}{k!(n-k)!}\frac{\partial^{n-k}}{\partial
z^{n-k} }\left\{\exp[z t]\left[\hat{K}(z)\right]^{n}_{z=\imath a
}\right\}\right]=
\]

\begin{equation}\label{sdp4}
=\textrm{Im} \left[\sum_{n=0}^{\infty}\frac{(2\imath a
\varepsilon)^{n}}{a\left(2a\right)^{2n}}\frac{\exp[\imath a
t]}{n!}
\sum_{k=0}^{n}\frac{(n+k)!}{k!(n-k)!}\left[t+\frac{\partial}{\partial
z }\right]^{n-k}\left[\hat{K}(z)\right]^{n}_{z=\imath a }\right].
\end{equation}
Defining the function~$$
\phi_2(\varepsilon,t,a)=\sum_{n=0}^{\infty}\frac{(2\imath a
\varepsilon)^{n}}{a\left(2a\right)^{2n}}\frac{1}{n!}
\sum_{k=0}^{n}\frac{(n+k)!}{k!(n-k)!}\left[t+\frac{\partial}{\partial
z }\right]^{n-k}\left[\hat{K}(z)\right]^{n}_{z=\imath a }$$ we may
write Eq.~(\ref{sdp4}) in a more concise way

$$F(t)=\textrm{Im} \left[ \exp[\imath a
t]\phi_2(\varepsilon,t,a)\right].$$ As before, a further
approximated expression for~$F(t)$ is given by

\begin{eqnarray}\nonumber
F(t)&\approx &\textrm{Im} \left[\exp[\imath a
t]\sum_{n=0}^{\infty}\frac{(2\imath a \varepsilon t
)^{n}}{a\left(2a\right)^{2n}n!}
\left[1+\frac{1}{t}\frac{\partial}{\partial z
}\right]^{n}\left[\hat{K}(z)\right]^{n}_{z=\imath a
}\right]\approx\\ \label{sdp5}&\approx &\textrm{Im}
\left[\exp[\imath a t]\sum_{n=0}^{\infty}\frac{(2\imath a
\varepsilon t )^{n}}{a\left(2a\right)^{2n}n!}
\left[\hat{K}(\frac{n}{t}+\imath a)\right]^{n}\right].
\end{eqnarray}
Finally, when~$\varepsilon$ is small enough, we rediscover the
naive solution

\begin{eqnarray}\nonumber
F(t)&\approx &\textrm{Im} \left[\frac{\exp\left[\imath
a\left(1+\varepsilon \frac{K(\imath a)}{2 a ^2}\right)t\right]}{ a
}\right]=\\ \label{sdp6}&=&\frac{\exp\left[\imath
a\left(1+\varepsilon \frac{\hat{K}(\imath a)}{2 a
^2}\right)t\right]-\exp\left[-\imath a\left(1+\varepsilon
\frac{\hat{K}(-\imath a)}{2 a ^2}\right)t\right]}{ 2\imath a }
\end{eqnarray}
that basically coincides with \eq{lapl6}. As for
$\phi_1(\varepsilon,t,z)$, the function~$\phi_2(\varepsilon,t,a)$
does not merely represent an exponential correction to the
exponential unperturbed solution. As previously stated, it is more
complex. Note that the exponential correction holds true only for
a sufficiently small~$\varepsilon$.

\section{Physical applications: the convoluted Lindblad equation}
\label{sec6} A wide range of applications exists for first and
second order integro-differential equations. In this section we
shall study a case that is well known in the scientific literature
for the quantum density matrix, the celebrated Lindblad
equation~\cite{lin}. We begin with

\begin{equation}\label{pa}
\frac{
\partial}{\partial t}\rho(t)=
-\frac{\imath}{\hbar}[H,\rho(t)]-L_D\rho(t)
\end{equation}
where~$L_D$ is a positive semidefinite Lindblad operator of the
form~\cite{gor}
$$L_D\rho=\frac{1}{\tau_D}[q,[q,\rho]]$$
with $q$ being the system's variable measured by the environment,
while $\tau_D$ represents the time scale of the
environment-induced measurement. The Hamiltonian $H$ and the
observable $q$ are properties of the system of interest, and they
are operators, as prescribed by quantum mechanics. To reduce the
number of parameters, we will set $\hbar=1$. For systems driven by
a time independent Hamiltonian, \eq{pa} can be promptly solved

\begin{equation}\label{pa_a}
\rho(t)=\exp[L t]\rho(0)=\sum_{k=1}^{n}c_k \exp[\lambda_k t]\rho_k
\end{equation}
where~$L=-\imath [H,\cdot]-L_D$ is the total liouvillian,~$\rho_k$
are its eigenvectors and~$\lambda_k$ its eigenvalues.  If the
matrix is an infinite dimensional matrix the sum can be extended
to the infinite. A natural generalization of \eq{pa} is

\begin{equation}\label{pa1}
\frac{\partial}{\partial
t}\rho(t)=\int_{0}^{t}\Phi(t-t')L\rho(t')dt'.
\end{equation}
Barnett and Stenholm~\cite{barnett} studied the above equation
considering the system as a harmonic oscillator embedded in a
reservoir. They showed that, even utilizing a simple exponential
kernel, the density matrix is positive definite only for a short
time. Wilkie previously~\cite{wil} showed that adding to \eq{pa1}
an inhomogeneous term with well defined characteristic, the
positivity properties of~$\rho(t)$ are preserved. Recently Bologna
\textit{et al.}~\cite{noi} demonstrated that it is always possible
to build a positive definite density matrix using the discrete
version of the Lindblad equation as a starting point. The passage
from natural time $n$ to continuous time $t$ is obtained
performing the subordination of the density matrix derived from
the discrete equation.

Starting from \eq{pa1} it is still an open problem to find
conditions on the kernel~$\Phi(t)$ that generate a positive
definite matrix. To achieve this, we rewrite the kernel operator
making a dimensionless parameter~$\varepsilon$ explicit, so that
$\Phi(t-t')L\to \varepsilon\Phi(t-t')L$. Then we evaluate the
Laplace transform of \eq{pa1}

\begin{equation}\label{pa2}
\hat{\rho}(u) =\left[u-
\varepsilon\hat{\Phi}(u)L\right]^{-1}\rho(0)
\end{equation}
where~$\hat{\Phi}(u)$ is the Laplace transform of~$\Phi(t)$. When
the inverse of the operator~$u- \varepsilon\hat{\Phi}(u)L$ acts on
the density matrix it can be written in terms of the eigenvalues
of $L$, namely~$$\frac{1}{u-
\varepsilon\hat{\Phi}(u)L}\rho(0)=\sum_{k=1}^{n}\frac{c_k}{u-
\varepsilon\hat{\Phi}(u)\lambda_k}\rho_k.$$ Each addend of the sum
is a Laplace transform of the form of \eq{fdp1} with $z=0$ and the
developing parameter parameter~$\bar{\varepsilon}_k\equiv
\varepsilon\lambda_k$. Note that $\bar{\varepsilon}_k$ is a linear
function of the eigenvalues. In principle, using the results of
Sec.~\ref{sec5}, and when~$\varepsilon$ is small enough, we can
write the inverse Laplace transform as

\begin{equation}\label{pa3}
\rho(t) =\sum_{k=1}^{n}c_k
\exp[\varepsilon\lambda_k\hat{\Phi}(0)t]\rho_k.
\end{equation}
As proven in Ref.~\cite{gor}, the density matrix given by
\eq{pa_a} has the proper physical meaning as does the matrix given
by \eq{pa3} since the only difference between the two is the time
scale factor $\varepsilon\hat{\Phi}(0)$. But, the application of
the method developed in Sec.~\ref{sec5} has to be carefully
considered. As stressed several times,~$\varepsilon$ has to be
small enough, one must count the factors in front of it, in order
to apply the naive solution. More precisely the following
sufficient conditions apply:
\newline i)~$\mid\!\!\hat{\Phi}(0)\!\!\mid<\infty$.
\newline ii)~$\hat{\Phi}(0)>0$.
\newline iii)~$\varepsilon\mid\!\!\lambda_k\!\!\mid\ll 1\,\,\textrm{for}\,\,k=1,2,\cdots$.
\newline The reasons for the above conditions rest on the
following arguments: Condition (i) is due to the requirement that
in general ~$\hat{\Phi}(z)$ has to exist and
consequently~$\hat{\Phi}(0)$ has to exist (see discussion in
Sec.~\ref{sec5}). Condition (ii) is required to preserve the sign
at the exponent in the exponential function [compare \eq{pa_a}
with \eq{pa3}]. Condition (iii) ensures a small developing
parameter. To insure a faster numerical convergence, condition
(iii) may be substituted with the more convenient one\newline
iv)~$\varepsilon \hat{\Phi}(0) \mid\!\!\lambda_k\!\!\mid\ll
1\,\,\textrm{for}\,\,k=1,2,\cdots$.\newline Considering the case
studied in Ref.~\cite{barnett} and applying the criteria of
condition (iii), we should select a value of~$\varepsilon$
sufficiently small such
that~$\varepsilon\!\mid\!\!\lambda_k\!\!\mid\ll 1$~$\forall k$.
The results of Ref.~\cite{barnett} show that the eigenvalues are
linearly divergent as a function of the index, viz.~$\lambda_k\sim
k$, so that condition (iii),
$\varepsilon\mid\!\!\lambda_k\!\!\mid\ll 1$, can never be
satisfied and the naive solution is not applicable. The density
matrix evaluated in Ref.~\cite{barnett} is not a positive definite
matrix. We conclude that this is due to the divergent structure of
its eigenvalues.

\begin{figure}[ht]
\begin{center}
\begin{minipage}[t]{0.3\linewidth}
\centering
\includegraphics[width=3.5cm, height=3.5cm,angle=0]{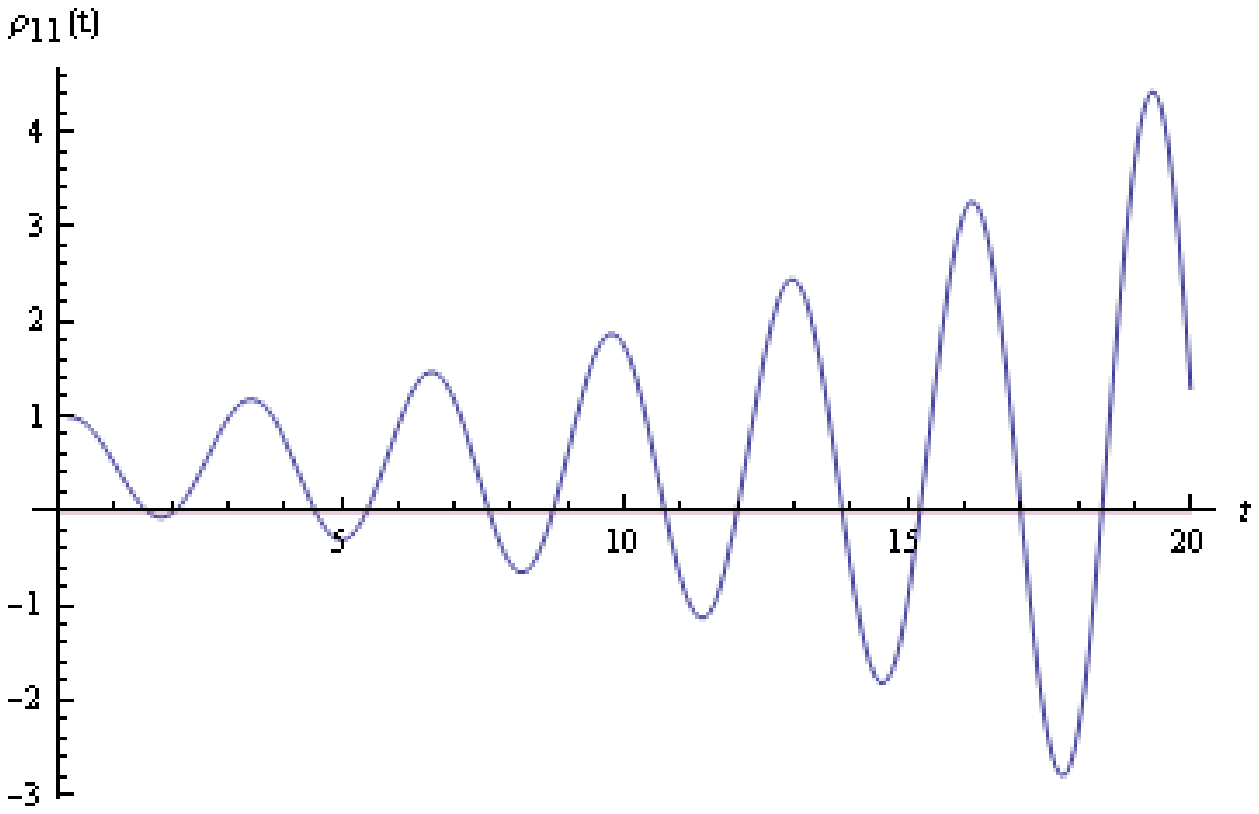}
\end{minipage}%
\begin{minipage}[t]{.3\linewidth}
\centering
\includegraphics[width=3.5cm, height=3.5cm,angle=0]{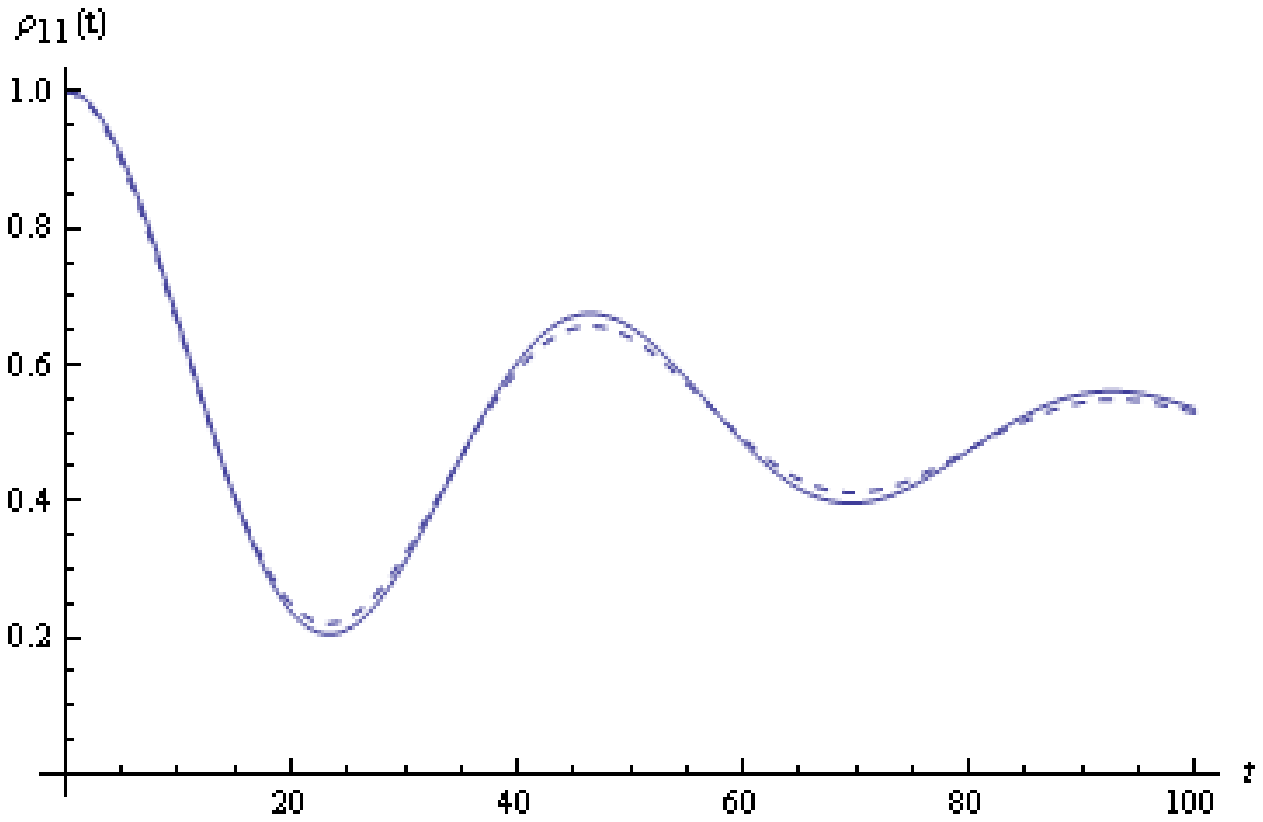}
\end{minipage}
\end{center}
\caption{\label{figfin}Left graphic: The plot of~$\rho_{11}(t)$,
Eq. (\ref{two1}). The value of the parameters are:~$
\epsilon=1.5,\,\,\gamma =8,\,\,\tau =1$ and~$\omega=5.5$ so that~$
2\varepsilon\omega/\gamma\approx 2.1$. Right graphic: The
continuous line represents the plot of~$\rho_{11}(t)$, Eq.
(\ref{two1}), while the dashed line represents the plot of the
approximate solution, Eq. (\ref{pa3}). The value of the parameters
are:~$\epsilon =0.1,\,\,\gamma=8,\,\,\tau=1$ and~$\omega=5.5$ so
that~$ 2\varepsilon\omega/\gamma\approx 0.14$.}
\end{figure}

To elucidate this last point, we can consider a finite-dimensional
density matrix where the condition $\mid\!\!\lambda_k\!\!\mid\leq
M$ $\forall k$, with $M$ as a finite positive number, can be
satisfied. The simplest case is a~$2\times 2$ density matrix.
Strictly speaking, to keep the analogy with the infinite
dimensional case studied in Ref.~\cite{barnett} we should consider
the convolution of the following equation

\begin{equation}\label{limb_simpl}
\frac{\partial}{\partial
t}\rho(t)=-\frac{1}{\tau}[\sigma_x,[\sigma_x,\rho(t)]]
\end{equation}
where the operator $q$ has been identified with Pauli's
matrix~$\sigma_x$. It is straightforward to show that the
convolution of the right side of \eq{limb_simpl} with an
exponential, always produces a positive definite density matrix.
We shall test the results of Sec.~\ref{sec5} considering the
convolution of the equation containing the full liouvillian used
in Ref.~\cite{noi},

\begin{equation}\label{jcp}
\frac{\partial}{\partial t}\rho(t)=L\rho(t)= -
\imath\omega[\sigma_x,\rho(t)]-\frac{1}{\tau}[\sigma_z,[\sigma_z,\rho(t)]]
\end{equation}
where~$\sigma_x$ and~$\sigma_z$ are Pauli's matrices. The
convoluted version of \eq{jcp} may be written as

\begin{equation}\label{jcp2}
\frac{\partial}{\partial
t}\rho(t)=\varepsilon\int_{0}^{t}\Phi(t-t')L\rho(t')dt=\varepsilon
\int_{0}^{t}\exp[-\gamma(t-t')]L\rho(t')dt
\end{equation}
where~$\varepsilon$ is a dimensionless parameter. The above
equation admits an analytical solution. We shall focus on the
diagonal element~$\rho_{11}(t)$ since~$\rho_{22}(t)$ can be
derived via the relation~$\rho_{11}(t)+\rho_{22}(t)=1$. Taking
into account that the eigenvalues of the liouvillian operator~$L$
are

$$\lambda_1=-\frac{4}{\tau },\,\,\,\,
\lambda_2=\frac{2 \left(-1-\sqrt{1-\tau ^2 \omega ^2}\right)}{\tau
},\,\,\,\,\lambda_3=\frac{2 \left(-1+\sqrt{1-\tau ^2 \omega
^2}\right)}{\tau },\,\,\,\,\lambda_4=0,$$ and assuming as the
initial condition~$\rho_{11}(0)=1$, we have

\begin{eqnarray}\nonumber
\rho_{11}(t)&=&\frac{1}{2}+\frac{e^{-\frac{t \gamma
}{2}}}{4\Gamma_1}\left(1-\frac{1}{\sqrt{1-\tau ^2 \omega
^2}}\right)\left(\Gamma_1 \cosh\left[\frac{t \Gamma_1}{2 \tau
}\right]+\gamma \tau \sinh\left[\frac{t \Gamma_1}{2 \tau
}\right]\right)+
\\
\label{two1}&+&\frac{e^{-\frac{t \gamma }{2}}}{4
\Gamma_2}\left(1+\frac{1}{\sqrt{1-\tau ^2 \omega
^2}}\right)\left(\Gamma_2 \cosh\left[\frac{t \Gamma_2}{2 \tau
}\right]+\gamma \tau \sinh\left[\frac{t \Gamma_2}{2 \tau
}\right]\right)
\end{eqnarray}
where we defined the quantities~$\Gamma_{1}$ and~$\Gamma_{2}$ as

$$\Gamma_1=\sqrt{\tau \left(-8 \varepsilon+\gamma ^2 \tau
-8\varepsilon\sqrt{1-\tau ^2 \omega ^2}\right)},\,\,\,\,
\Gamma_2=\sqrt{\tau \left(-8\varepsilon+\gamma ^2 \tau +
8\varepsilon\sqrt{1-\tau ^2 \omega ^2}\right)}.$$ From \eq{two1}
we can deduce that, in general,~$\rho_{11}(t)$ is neither a
positive quantity $\forall t$ nor a quantity smaller than the unit
$\forall t$. Applying the method developed in Sec.~\ref{sec5}, the
analysis of the eigenvalues shows that there are two cases of
interest:~$\tau\omega <1$ corresponding to real eigenvalues,
and~$\tau\omega>1$ corresponding to complex eigenvalues. To
simplify the two cases, consider~$\tau\omega \ll 1$
and~$\tau\omega \gg 1$. In the first case, condition (iv) may be
written as

\begin{equation}\label{cond1}
\frac{4\varepsilon}{\gamma\tau}\ll 1
\end{equation}
whereas in the second case it may be written as
\begin{equation}\label{cond2}
\frac{2\varepsilon\omega}{\gamma}\ll 1.
\end{equation}
Fig.~\ref{figfin} shows the plot of~$\rho_{11}(t)$ for given
values of the parameters
$\varepsilon,\,\,\omega,\,\,\tau,\,\,\gamma$. In the left graphic,
condition~(\ref{cond2}) is violated because~$
2\varepsilon\omega/\gamma\approx 2.1$, and~$\rho_{11}(t)$ does not
have physical meaning. In the right graphic, on the contrary,
condition~(\ref{cond2}) is satisfied as~$
2\varepsilon\omega/\gamma\approx 0.14$, and~$\rho_{11}(t)$ is a
positive function smaller then the unit~$\forall t$.

\section{Conclusions}
This paper introduced a procedure to obtain an analytical
asymptotic expression for the solution of first and second order
integro-differential equations containing an arbitrary kernel. We
found an asymptotic expression for the correspondent inverse
Laplace and Fourier transforms containing an arbitrary
function~$\hat{K}(u)$. It was shown that a first order
integro-differential equation is asymptotically driven by an
equation that is independent from the specific form of the kernel
of the integro-differential equation. A general expression for the
desired asymptotic solution was also given. This result was
applied to the convoluted version of the Lindblad equation
explaining why even a simple kernel, such as an exponential
function, does not generate a positive definite density matrix.
Sufficient conditions on the kernel function so as to generate a
positive definite density matrix were given at the end of
Sec.~\ref{sec6}. Indeed, these conditions showed that the
structure of the eigenvalues of the liouvillian operator plays a
crucial role in determining the positivity of the density matrix.

\section*{Acknowledgments}
The author would like to thank Catherine Beeker for her editorial
contribution.

\end{document}